# Low-Dimensional Conduction Mechanisms in Highly-Conductive and Transparent Conjugated Polymers

Asli Ugur[1,2], Ferhat Katmis[3], Mingda Li[4], Lijun Wu[5], Yimei Zhu[5], Kripa K. Varanasi*[,2], Karen K. Gleason*[,1]

[1]Department of Chemical Engineering,
Massachusetts Institute of Technology, Cambridge, Massachusetts 02139, USA
[2]Department of Mechanical Engineering,
Massachusetts Institute of Technology, Cambridge, Massachusetts 02139, USA
[3]Department of Physics,
Massachusetts Institute of Technology, Cambridge, Massachusetts 02139, USA
[4]Department of Nuclear Science and Engineering,
Massachusetts Institute of Technology, Cambridge, Massachusetts 02139, USA
[5]Condensed Matter Physics & Materials Science Department,
Brookhaven National Laboratory, Upton, NY 11973, USA

Charge transport mechanisms in conjugated polymers are widespread interest for enabling high-performance and low-cost optoelectronic and thermoelectric devices.[1] Optimal device performance for high-end technologies requires a fundamental grasp of transport properties at the nanoscale.[2] However, the understanding of conduction mechanisms remains incomplete even after several decades of intensive investigation.[3] Recently clues have emerged which shed light on complicated microscale conduction pathways.[4] These exciting finders further the debate on charge transport in conducting polymers, which are neither fully amorphous nor completely crystalline. There is a strong tendency for collective charge motion along the ordering direction of the polymer chains where strong covalent bonds form along the same direction.[5] In addition, strong backbone, length of the conjugation chain and most importantly polymer crystallinity have important effects on conduction. Charge-carrier mobility depends not only on polymer crystallinity and crystallite orientation at nanoscale, but also on the connectivity of macroscopic domains and morphological imperfections.[5] Among the conjugated polymers, poly (3,4-ethylenedioxythiophene) (PEDOT) displays a valuable combination of high electrical conductivity and stability. PEDOT is desired for commercial applications including photodiodes, capacitors, sensors, and organic light-emitting diodes,[6] and is also a promising candidate for integration into future macro- and thermo-electronic devices.[7, 8] The desirable features of PEDOT derive from the chemical structure of the 3,4-ethylenedioxythiophene monomer.[9] The backbonded ring at 3,4-positions of the 5-member thiophene ring allows polymerization to occur only through the 2,5- positions. The resulting conjugated PEDOT chains have a linear structure and hence a propensity to crystallize. In addition, the pair of electronegative oxygen atoms in the backbonded ring stabilizes the presence of the cationic doping species required for conductivity. Previously, we have reported oxidative chemical vapor deposition (oCVD) as a means of synthesizing conductive PEDOT layers.[10] Additionally, we demonstrated the ability to graft PEDOT directly to its substrate. The covalent chemical bonds formed across the interface eliminate cracking and delamination; which facilitates well-defined lithographic patterns to resolved in PEDOT for feature size as small as 60 nm.[11]

In the current work, the oCVD synthesis of PEDOT is utilized to control crystallite size as well as the orientation of these ordered domains. Significant increases in crystallite sizes are achieved by increasing deposition temperature. Additionally, we show for the first time that oCVD grafting produce transverse alignment of polymer chains in the crystalline domains in the near interface region. These uniquely structured polymer films allows us to quantitatively access the predictions of different charge transport models. We show that, the electrical conductivity increases with increasing crystallite sizes and interestingly, with decreasing thickness in contrast to inorganic materials. For these observations a linear combination of Mott`s and Efros-Shklovskii`s variable range hopping (ES-VRH) models is used that provides an explanation for the dimensionality of the films and the inter-crystallite hopping. Based on the excellent agreement between direct experimental evidence and our model, we hypothesize that high overlap of the total wavefunctions of the ground state of the charges at each crystalline domain, which increases with increased crystal domain size; is the main reason for the increased conductivity. For ultrathin films, the conductivity can be well described by Mott`s VRH for highly ordered structures. However, the ES-VRH model, which applies to disordered systems,[12] is needed to quantitatively predict the data for thicker films. Increased disorder results in 3D conduction pathways that diminish the charge transport; in contrast, less disordered (2D) utra-thin films have lower resistances since the conduction pathways are better aligned. In the future, we anticipate the ability to differentially influence the transport of phonons and charge carriers across grafted versus ungrafted interfaces as a key means by which thermoelectric interfaces employing PEDOT can be fundamentally understood and optimized.

The grafted and ungrafted oCVD PEDOT thin films were grown on Si(100) having a native oxide layer (experimental details can be found in the supplementary online materials (SOM)). oCVD is a one step process in which both monomer and oxidant ($FeCl_3$) are delivered in the vapor phase in a vacuum. Silicon substrates are exposed to 3,4-ethylenedioxythiophene monomer delivered from a monomer jar from a side of the vacuum chamber. The substrate temperature varied from 100 to 200 °C under ~ 0.1 mTorr chamber pressure. For all depositions the flow rate of PEDOT is kept constant at ~ 5 sccm and deposition time varying from 15 to 60 min. To graft PEDOT, a silane coupling agent is used to form a robust bonding transition from inorganic to organic material. The (100) oriented silicon substrate with native silicon oxide layer is treated with oxygen plasma (29.6 watts, 15 min) to form hydroxyl groups. The sample is then exposed to trichlorovinylsilane (TCVS – $C_2H_3SiCl_3$) to form a vinyl terminated surface to which PEDOT can graft.

X-ray diffraction (XRD) measurements were carried out using $CuK_{\alpha 1}$ radiation on both grafted and ungrafted films. Large area XRD mapping further reveals the distribution of crystallite orientations through two types of reflections, ($h$00) and (0$k$0). For the ungrafted samples (**Figure 1b**), the predominant reflection is (0$k$0), corresponding to the PEDOT backbone aligned parallel surface and packed together with the π-π stacking direction, which is perpendicular (**Figure 1a**(ii)) to the plane of the substrate. This orientation is known as face-on packing. In contrast, for the grafted film, both (0$k$0) and ($h$00) diffraction families are observed (**Figure 1c**). The presence of the ($h$00) reflection is clear evidence that grafting creates a new population of crystalline domains in which the backbone of the PEDOT chain is oriented, likewise, parallel to the surface, however, with the π-π stacking parallel to the plane of the substrate surface (**Figure 1a**(i)). This oriention is known as edge-on packing. Existence of vinyl bonds at the interface, which are created by pre-treatment of the substrate, changes the surface energy and forces to change the bond type from weak van der Waals bonds to strong covalent bonds at the interface. These created bonds might force the backbone of the polymer to pack in a different orientation. Grafted orientation represents ~79% of all oriented domains in this 180 nm thick film, which is calculated by using Scherrer equation.[13] We hypothesize that the grafted chains are in the near interface region, since the grafting starts at the interface where viny bonds are created and that thinner films would display a higher fraction of the grafted orientation. There is also a small fraction of perpendicular orientated backbone to the substrate surface (**Figure 1a**(iii)) for grafted films, which is not observable with XRD, however these domains are visible in transmission electron microscopy images on grafted samples (see SOM **Figure S7**).

Choosing between grafted and ungrafted oCVD PEDOT offers bottom-up controllability of the polymer films, which also provides control of the electrical properties. Here the highest in-plane conductivity values are obtained for the ungrafted films (**Figure 4**) for which the XRD reveals only (0$k$0)–oriented crystallites. In these crystallites, the conjugated PEDOT backbone is aligned parallel to the substrate (**Figure 1**), where the π-π stacking is perpendicular to the substrate surface which is the densest packing configuration and hence the most efficient transport pathway. Thus, the conductivity results demonstrate that the crystallization and alignment of the PEDOT chains along the substrate on the nanoscale improve the macroscopically observed charge transport along the same direction. The orientation of a single chain determines the contribution of polarons, while the order packing of multiple aligned chains influences the contribution of the bipolarons.[8]

It is important to realize if domain boundaries have lower electrostatic potential than the crystalline domains, and the domains themselves have been hypothesized to be charge trapping centers.[14] Mean crystallite sizes dervived from the Scherrer equation[13] from the observed XRD linewidths are 3 nm for (0$k$0) and 5.5 nm for ($h$00)-type oriented crystallites. If scattering from the crystallites and their domain boundaries dominated the charge transport in the PEDOT films, we would expect the ungrafted film which is comprised entirely of the smaller (0$k$0) oriented crystals to have a reduced conductivity relative to its grafted counterpart. However the opposite is observed (**Figure 4**), suggesting that another mechanism dominates the charge transport.

To better understand the role of the nanoscale crystallinity on electrical conductivity, advanced electron microscopy measurements were undertaken to reveal the distribution of crystallite sizes present in the oCVD PEDOT films. These measurements complement the XRD results and provide direct proof of the well-controlled chain alignment. Trends in the distribution of domain sizes with variation in substrate temperature in the grafted films were found to be similar to their ungrafted counterparts. The nanocrystalline films contain highly ordered regions separated by disordered (amorphous) chains (see schematic illustration in SOM as **Figure S3**) and effects of temperature on formation of such domains are crucial. The existence of different size of domains is confirmed by the high angle annular dark field (HAADF) images with scanning transmission electron microscopy (STEM), where image contrast is roughly proportional to the square of the atomic number of probing chemical species (the HAADF images are shown in **Figure 2**). The HAADF images of the grafted PEDOT films grown at 100 °C and 200 °C substrate temperatures are shown in **Figure 2a** and **b**, respectively. The left region is the substrate and $SiO_2$, while the pink region on the right is the Pt protection layer. The PEDOT thin film is sandwiched in the middle (dark blue). In order to verify the domain distribution and sizes of the crystalline PEDOT inside the thin film, high-resolution HAADF images are obtained (**Figure 2c** and **d**) in STEM mode. The crystalline structures in the PEDOT are clearly seen in the magnified images **Figure 2c** and **d**.

Both the size distribution of the domains and their inter-domain spaces change with deposition temperatures (**Figure 2e**). At low temperature, the domain sizes are small (~2 nm) as shown by a single peak in the histogram. At elevated temperature, the scenario changes dramatically and leads to significant increase in conductivity (**Figure 4a-b**). As seen in the **Figure 2b-d**, the crystalline sizes dramatically increase, up to 10 nm, indicating either by direct growth of larger domains or coalescence of small domains induced by the increased mobility of the polymer chains at elevated temperatures. The later hypothesis is consistant with some small domains remaining in regions between the larger domains, which are indeed observed in the histogram.

The remarkable ordering of PEDOT crystallite sites resulting from oCVD synthesis may facilitate hopping of electrons between the crystalline domains. In order to further verify the hopping conduction mechanism, we performed temperature-dependent electrical transport measurements and analyzed these results using a "coarse–grained" variable range hopping (VRH) model. The original VRH model describes the low-temperature conductivity caused by the hopping from localized electrons[15] from one atomic site to another. In the present situation, instead of atomic sites, intra-crystallite conduction electrons (i.e. extended electron clouds) are considered localized when hopping among crystallites (inter-crystallite hopping). This hopping contributes to the total electrical conductivity (**Figure 3a**). Therefore, the hopping is no longer from one atomic site to its neighbor atoms, but "coarse-grained" from one crystallite site to another. When the crystallites are larger, the overlap between their total wavefunctions increases, leading to improved transport pathways in which the resistive amorphous areas no longer dominate the overall conductivity. The Mott's VRH applies to the condition where Coulomb interactions between the localized electrons are negligible. When inter-crystallite long-range

Coulomb interactions becomes significant, these interactions open up a soft Coulomb gap, as described by the Efros-Shklovskii (ES) type of VRH.[16] The total normalized resistivity can be empirically treated as a linear combination of Mott's VRH and ES-VRH:

$$\frac{\rho(T_{eff})}{\rho(300K)} = \left( \beta_1 \exp\left[ -(T_M/T)^{\frac{1}{1+d}} \right] + \beta_2 \exp\left[ -(T_{ES}/T)^{\frac{1}{2}} \right] \right)^{-1}, \quad (1)$$

where $\beta_1$ and $\beta_2$ are dimensionless amplitudes for the Mott and ES type of conductivity, respectively, and their ratio determines whether Coulomb interaction is significant or not. The samples that are less disordered can overcome these interactions and the conduction can be explained purely with Mott`s VRH. On the other hand, with increased disorder the ES-VRH starts to arise. $T_M$ is Mott's transition temperature for the onset of conduction, and $T_{ES}$ is the ES type VRH transition temperature which satisfies $T_{ES} \propto 1/\varepsilon$, where ε is the dielectric constant. In the first term of Eq. 1, the exponent, $d$, is the effective dimensionality for the Mott-type conduction which determines the dimension of the film

The "coarse-grained" VHR model (Eq. 1) provides a superb description of the observed temperature dependence of resistivity from room temperature down to 10 K, measured using the conventional Van der Pauw geometry (**Figure 3b**). The value at 0 K is obtained by spline extrapolation (**Figure 3b**, inset). Semiconductor-like behavior, d$\rho$/d$T$, is observed for all samples. Since the ungrafted samples display both the highest conductivies and only a single crystal orientation, we selected only the the ungrafted for measurement of the temperature dependent transport characteristics. A non-linear least squares regression of the observed change in conductivity with temperature (**Figure S5** in SOM) for each of the 3 samples measured is

A. UGUR et al.

performed against Eq. (1) by forcing sample #1 to be 3D, sample #2 to be 2D, and sample #3 to be 2D, as tabulated in **Tab. S3** in SOM. Sample #1, which has the lowest resistivity, is deposited at 200 °C with conductivity of 2050 S/cm (52 nm), Sample #2 has an intermediate resistivity which is also deposited at 200 °C but is thicker (124 nm) than sample #1 with conductivity of 885 S/cm, and Sample #3 has the highest resistivity with lower preparation temperature, 100 °C with conductivity of 320 S/cm (145 nm). All three samples were post rinsed with 5 mol L$^{-1}$ HBr acid to remove the excess oxidant and dope films with Br- counter-anions to obtain low resistivity layers. From the table we see directly when we fix the incorrect value for dimensionality and let the nonlinear optimization runs, the residual (sum of mean square error) increases significantly. For sample #1, the thinnest of the set, $d = 2$ (2D), in contrast, both thicker samples, #2 and #3, $d = 3$ (3D) provides the statistically best regression to the data, as shown in **Figure 3b** and table in **Tab. S3** in SOM. Samples #1, the thinnest among the samples, displays good agreement with pure Mott's VRH, (e.g. $\beta_2 = 0$ in Eq. 1; see SOM for detail) where there is no contribution of ES-VRH. However, it can be seen in **Figure S4** in SOM that the fitting has some deviations for all samples, when only Mott's VRH is used. Therefore, ES-VRH is also included where the fittings become excellent as shown in **Figure 3b**. Even though sample #1 and sample #2 are prepared with the same deposition temperature (200 °C) due to increased disorder in the thicker films, explained by ES-VRH, the charge transport pathways become 3D, which increases the resistivity of the films. The same effect is also observed more extensively for sample #3 that is deposited at the lower substrate temperature of 100 °C, with comparable thickness to sample #2. This sample has smaller crystallites as shown in **Figure 2a** compared to the films deposited at higher temperatures in **Figure 2b**. A remarkably good match is obtained between the resistivity data and Eq. 1 (Mott+EH VRH); for sample #2 $\beta_2 = 6.74$, and for sample

#3, $\beta_2$ = 116.2, which indicate a ~ 77.5% and 96.8% contribution of the ES-VRH conduction model, respectively, representing significant inter-crystallite Coulomb interaction. For sample #1, $\beta_2$ = 0.69 indicating the weakest contribution of ES-VRH (46%). In sample #3, the origin of the weaker Coulomb screening effect is consistent with a higher fraction of amorphous regions, since amorphous regions generally have smaller dielectric constant ε compared with crystalline regions. It would be expected to have the lowest dielectric constant and correspondingly the weakest Coulomb screening effect.

Moreover, since $T_M \propto (1/S)^3$ where $S$ is the average crystallite size, values of $T_M$ extracted from Eq. 1 correspond to an average crystalline size ratio S1:S2:S3 ~ 3:2:1. This ratio is consistent with the direct observation of crystallite size by HAADF, where the higher deposition temperature leads to larger crystallite size. Obtained crystalline size ratio coincides with the values of $T_{ES}$ obtained from Eq. 1. Sample #3 has largest $T_{ES}$, together with the largest increase of resistivity at low temperature (SOM). The room temperature conductivity for all samples reduces while decreasing the temperature. The most obvious trends observed here is that sample #3 shows much larger variation in resistivity near 0 K than the other two samples and that this observation is additional support of the contribution of the ES-VRH conduction mechanism in sample #3. To generalize a relation for 2D and 3D type samples, the reduction of resistivity as dimension $d$ decreases, at various sizes of domains, are generic. The logarithmic of normalized resistivity ratio $\log(\rho_{2D}/\rho_{3D}) \left( \frac{\rho_{2D}}{\rho_{3D}} \equiv \frac{\rho_{2D}(T)/\rho_{2D}(300\,\mathrm{K})}{\rho_{3D}(T)/\rho_{3D}(300\,\mathrm{K})} \right)$ between a 2D and 3D-like samples are shown in **Figure 3c**. If two samples have the same domain sizes, with identical $T_M$, the resistivity increases more rapidly in 3D sample. The parameters are taken from 2D-like sample #1 and 3D-

*A. UGUR et al.*

like sample #3, respectively, but keeping them with the same. It can be seen directly that $\rho_{2D} < \rho_{3D}$ is valid as long as T>30 K.

It is desirable to establish the limitations for controlling the conductivity within the variety of polymer films. For this purposes, we analyze the different parameters one-by-one hitherto for mostly between so called 3D- and 2D-like arrays of PEDOT crystallites. To further enlarge the hypothesis of 2D vs. 3D conduction, three sets of film thickness ranges were investigated; *i)* ultra-thin limit, less than 10 nm, *ii)* intermediate case of between 10 nm up to 80 nm, and *iii)* bulk-dominated phase where thicknesses are thicker than 100 nm. For films thinner than 80 nm, we hypothesize that the surface helps drive ordered assembly of the crystallites of the first crystallites formed. At each substrate temperature over the range (except for ultra-thin films) from 100 to 200 °C, four variants were prepared: grafted or ungrafted films which were either rinsed with methanol only or additionally rinsed with HBr. For the intermediate thicknesses films, for each of the 4 variants, the conductivity (σ) increased monotonically with substrate temperature (**Figure 4a**). The highest conductivities for the intermediate thickness range, σ ~ 2000 S/cm, is achieved for the ungrafted and HBr rinsed condition at a substrate temperature of 200 °C. The next highest values, σ ~ 1800 S/cm, are observed with grafted and HBr rinsed sample grown at the same substrate temperature. The HBr rinsed films generally have about two times higher conductivity than their comparable MeOH only rinsed counter parts. Pairwise comparison between the ungrafted and grafted films, which experienced the same rinsing treatment, reveals that grafting reduces conductivity in all but one case. To further probe the effect of thickness on conductivity, samples with less than 10 nm thicknesses were deposited at the 150, 175 and 200 °C substrate temperatures which were ungrafted and HBr rinsed. An increase in conductivity close to 3700 S/cm is obtained for high growth temperatures (presented in **Figure 4a** as star symbols), which is a clear indication that these nanoscale layers have improved charge transport properties. In **Figure 4b**, all the films measured were above 100 nm thick. All the same trends with substrate temperature, grafting and rinsing are observed in **Figure 4a** are found again in **Figure 4b**. Notably, in the thicker films of **Figure 4b**, the conductivities are lower than observed for their thinner counterparts shown in **Figure 4a**.

Next generation optoelectronic devices, such as displays, solar-cells or touch screens, require flexibility and processability over large areas. Currently, Indium tin oxide (ITO) is used as transparent electrode material extensively that has ideal optoelectronic properties. However, ITO suffers from high cost and brittleness. These two major drawbacks hinder the use of ITO in large-area flexible electronics for next generation devices. Transparent and conductive polymers are promising candidates that could replace ITO with their low cost and processability. The ultra-thin PEDOT films (<10 nm) that are studied in this work have excellent transparencies that reached up to ~ 98% (**Figure S6** in SOM) at 550 nm with sheet resistances down to 500 Ω/□. To analyze optical figure of merit for conducting transparent thin films, the following equation is used where transmittance ($T$) and sheet resistance ($R_s$) are linked;

$$T = (1 + \frac{Z_0}{2R_s} \frac{\sigma_{dc}}{\sigma_{op}}) \qquad (2)$$

where $Z_0$ is the impedance of free space (377 Ω). The dc conductivity to optical conductivity ratio ($\sigma_{dc}/\sigma_{op}$), increases with increasing deposition temperature and reaches up to ~35.3 for 200 °C for PEDOT films (**Figure 4c**) which is comparable with the literature values for PEDOT:PSS and sufficient for transparent electrode applications.[17]

In summary, the combination of multiple characterization methods and controlled synthesis by oCVD gives rise to deeper understanding of the charge transport properties of PEDOT polymer films. In ungrafted films, the PEDOT chains are primarily in face-on configuration. Crystallites with this same orientation were observed in grafted films; however, in addition, a significant fraction of crystallites also formed in the edge-on configuration. There is also a small fraction of crystallites, where the backbone is aligned perpendicular to the substrate for grafted films. However, the latter configuration is only observable with STEM. For both grafted and ungrafted films, increasing substrate temperature had the largest impact on increasing conductivity where the domain sizes are clearly visualized by STEM. Additionally, Br doping resulting from acid rinsing consistently decreases the sheet resistance for both grafted and ungrafted films. We successfully applied the coarse-grained variable range hopping theory, which describes the inter-crystallite electron hopping, to explain the resistivity as a function of temperature. Despite the extended electronic state within one crystallite, it is the overlap between the extended intra-crystallite electronic states which contributes to the final conductivity. Most importantly, the effect of dimensions (2D vs 3D) and the crystalline domain sizes were extracted from the model, and showed excellent agreement with both X-ray and electron diffraction based results. This paves the pathway for the new mechanism of conductivity enhancement, in addition to the conduction within polymer chain. Electrical conduction in conjugated polymeric nanolayers is crucial to the emerging technological interest for high performance optoelectronic and thermoelectric devices.

**Methods**

Detailed description of sample preparation for deposition by oCVD and STEM characterization, XRD measurement configuration and details about modelling is provided in supplementary online materials (SOM).

**Supporting Information**

Supporting Information is available from the Wiley Online Library or from the author.

**Acknowledgements**

The authors acknowledge financial support from the MIT Institute for Soldier Nanotechnologies (ISN) under Contract DAAD-19-02D-0002 with the U.S. Army Research Office. Part of this work was carried out at the CMSE shared experimental facilities, and we would like to thank S. Speakman for assistance and J. Moodera for fruitful discussions. The Work at BNL was supported by the U.S. Department of Energy, Office of Basic Energy Science, Material Science and Engineering Division, under Contract No. DE-AC02-98CH10886.

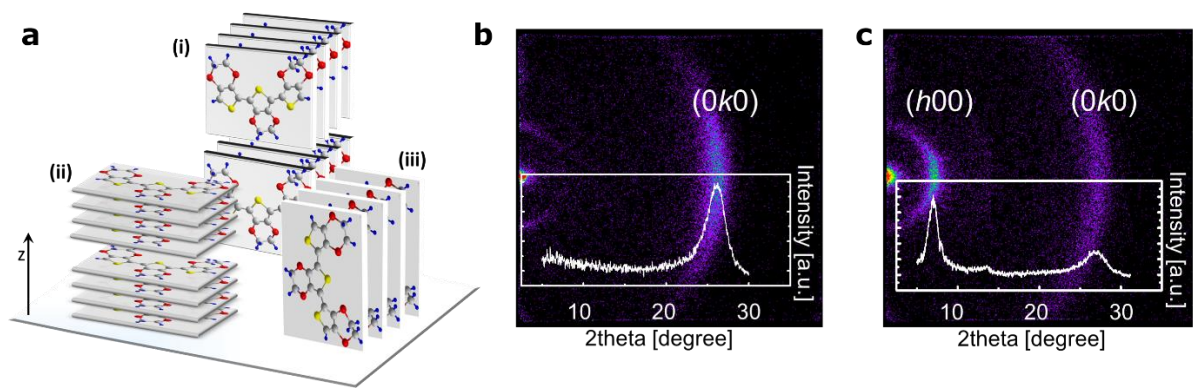

**Figure 1.** a) Three possible orientations for ordered PEDOT chains with respect to the substrate plane, two of which have their conjugated backbones parallel to the interface, (i) and (ii), while the orientation is perpendicular for (iii). The XRD maps (θ-2θ) for b) ungrafted and c) grafted oCVD PEDOT. Only (0*k*0) reflections appear for the ungrafted film, corresponding to the parallel orientation (ii). For the grafted film, both (*h*00) and (0*k*0) reflections are observed, indicating the presence of (i) and (ii) orientations.

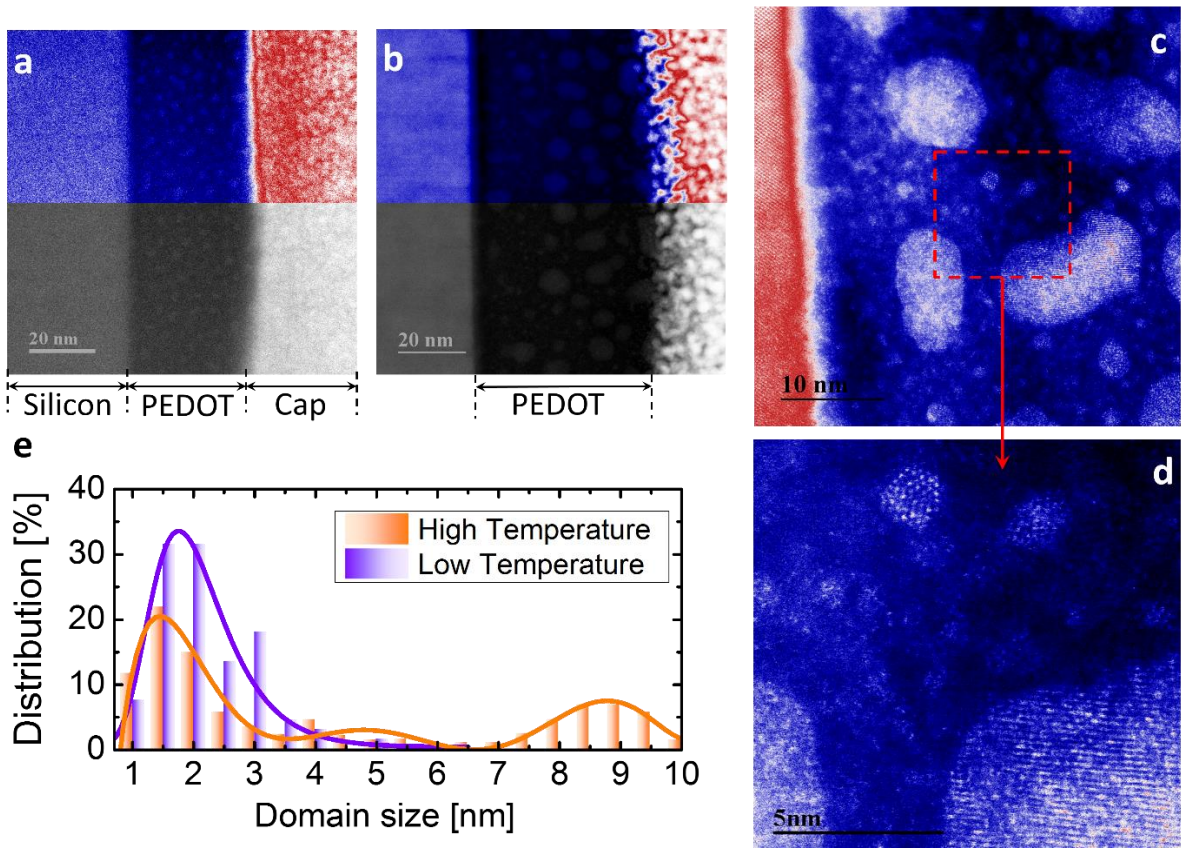

**Figure 2.** The High Angle Annular Dark Field (HAADF) Scanning Transmission Electron Microscopy (STEM) image of Cross sections of the interfaces of grafted PEDOT films grown at a) 100 °C and b) 200 °C. The top halves of both images are color-enhanced to clearly elucidate the crystalline domains surrounded by an amorphous matrix. High-resolution images for the film synthesized at 200 °C are shown in c) and enlarged in d), providing a direct evidence on the well-oriented large crystallites. e) Histogram of statistical domain size distribution obtained from images (a) and (b), showing the broader distribution and larger crystallite size for the film grown at high temperature (200 °C). STEM images indicate all three possible orientations for grafted samples as shown in Figure 1a (see SOM for detail).

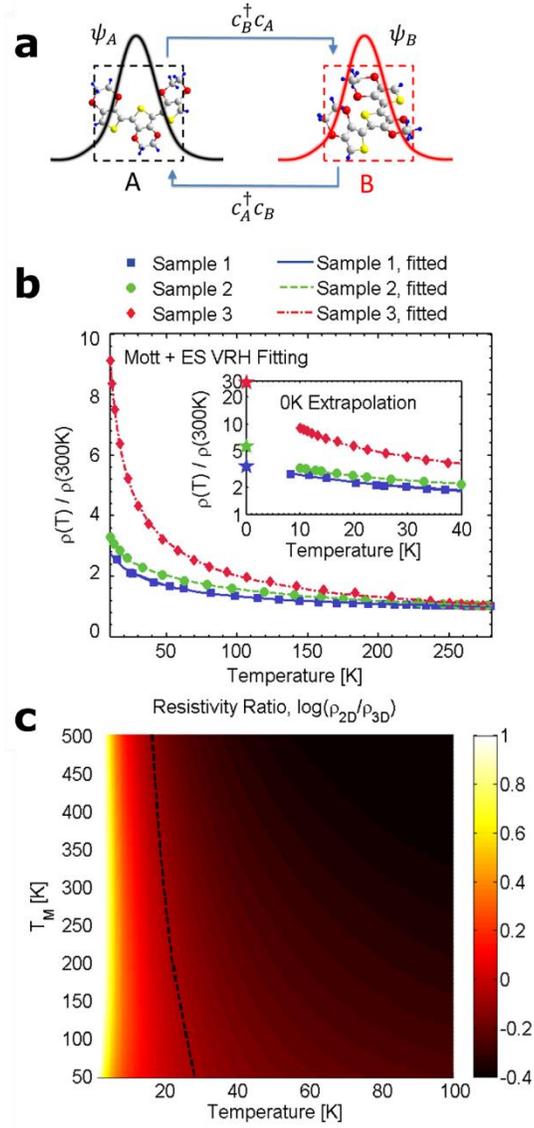

**Figure 3.** The measured conductivities and predicted values from the combined Mott + ES VRH model (Eq. 1). a) Microscopic representation of coarse-grain model for different domains is illustrated. Electron wave functions in crystalline domains A and B are denoted as $\psi_A$ and $\psi_B$, respectively, which are also separated by an amorphous region. The conductivity originates from hopping of the electrons. The hopping from $B$ to $A$ is depicted as $c_A^\dagger c_B$. b) Low temperature resistivity measurements for Sample #1 is grown at 200 °C in where 2D conduction is dominated (52 nm), sample #2 is grown at 200 °C where the 3D conduction is dominated (124

nm), and sample #3 is grown at 100 °C where 3D conduction is dominated (145 nm). Inset figure shows the normalized resistivity at low-temperature region with a highlight of 0 K extrapolation. c) Extended model represented as logarithmic of the normalized resistivity ratio $\log(\rho_{2D}/\rho_{3D})$ between a 2D-like sample and a 3D-like sample via varying the parameter $T_M$ and T. The black dot trend line separates the region greater than 0 (2D has larger resistivity) and smaller than 0 (2D has smaller resistivity).

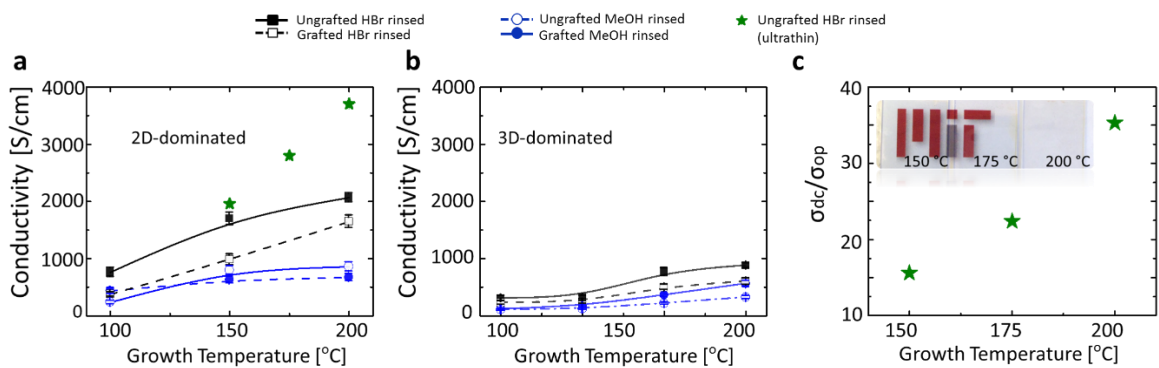

**Figure 4.** The magnitude of room temperature conductivity for different sample sets comprises 3D-like where amorphous region is dominated and/or 2D-like where crystallite region is dominated. a) Single step deposition of films < 80 nm thick (2D-dominated). The highest conductivity (star symbols) is for films <10 nm thick. b) Single-step deposition of films >100 nm thickness (3D conductivity). c) The dc conductivity to optical conductivity ratios ($\sigma_{dc}/\sigma_{op}$) of ultra-thin highly conductive PEDOT films (<10nm). Inset shows the highly transparent PEDOT films on glass slides.